\documentclass[aps,twocolumn,superscriptaddress,showpacs,amssymb,prb,floatfix,longbibliography]{revtex4-2}
\usepackage{graphicx }% Include figure files
\usepackage[]{hyperref}
\usepackage{amsmath}
\usepackage{amssymb}
\usepackage{siunitx}
\usepackage{bm}

\hypersetup{colorlinks=true,linkcolor=blue,citecolor=blue,urlcolor=blue,pdfpagemode=UseNone}

\newcommand{\slrr}      {$T_1^{-1}$}

\begin{document}

\title{Magnetic structure and Kondo lattice behavior in CeVGe$_3$: an NMR and neutron scattering study}
\author{C. Chaffey}
\affiliation{Department of Physics and Astronomy, University of California Davis, Davis, CA }
\author{H. C. Wu}
\email{wu.hung.cheng.d2@tohoku.ac.jp}
\affiliation{Institute of Multidisciplinary Research for Advanced Materials, Tohoku University, Sendai 980-8577, Japan}
\author{Hanshang Jin}
\affiliation{Department of Physics and Astronomy, University of California Davis, Davis, CA }
\author{P. Sherpa}
\affiliation{Department of Physics and Astronomy, University of California Davis, Davis, CA }
\author{Peter Klavins}
\affiliation{Department of Physics and Astronomy, University of California Davis, Davis, CA }
\author{M. Avdeev}
\affiliation{Australian Centre for Neutron Scattering, Australian Nuclear Science and Technology Organisation, Kirrawee DC, NSW 2232, Australia}
\affiliation{School of Chemistry, The University of Sydney,
Sydney 2006, Australia.}
\author{S. Aji}
\affiliation{Institute of Multidisciplinary Research for Advanced Materials, Tohoku University, Sendai 980-8577, Japan}
\author{R. Shimodate}
\affiliation{Institute of Multidisciplinary Research for Advanced Materials, Tohoku University, Sendai 980-8577, Japan}
\author{K. Nawa}
\affiliation{Institute of Multidisciplinary Research for Advanced Materials, Tohoku University, Sendai 980-8577, Japan}
\author{T. J. Sato}
\affiliation{Institute of Multidisciplinary Research for Advanced Materials, Tohoku University, Sendai 980-8577, Japan}
\author{V. Taufour}
\affiliation{Department of Physics and Astronomy, University of California Davis, Davis, CA }
\author{N. J. Curro}
\email{njcurro@ucdavis.edu}
\affiliation{Department of Physics and Astronomy, University of California Davis, Davis, CA }

\date{\today}
\begin{abstract}
We present nuclear magnetic resonance (NMR), neutron diffraction, magnetization, and transport measurements on a single crystal and powder of CeVGe$_3$.  This material exhibits heavy fermion behavior at low temperature, accompanied by antiferromagnetic (AFM) order below 5.8 K.  We find that the magnetic structure is incommensurate with AFM helical structure, characterized by a magnetic modulated propagation vector of $(0, 0, 0.49)$ with in-plane moments rotating around the $c$-axis. The NMR Knight shift and spin-lattice relaxation rate reveal a coherence temperature $T^*\sim 15$ K, and the presence of significant antiferromagnetic fluctuations reminiscent of the archetypical heavy fermion compound CeRhIn$_5$. We further identify a metamagnetic transition above $H_m\sim 2.5$ T for magnetic fields perpendicular to $c$.  We speculate that the magnetic structure in this field-induced phase consists of a superposition with both ferromagnetic and antiferromagnetic components, which is consistent with the NMR spectrum in this region of the phase diagram.  Our results thus indicate that CeVGe$_3$ is a hexagonal structure analog to tetragonal CeRhIn$_5$.

\end{abstract}

\maketitle
%%%%%%%%%%%%%%%%%%%%%%%%%%%%%%%%%%%%%%%%%%%%%%%%%%%%%%%%%%%%%%%%%%%%%%%%%%%%%%%%%%%%%%%%%%

\section{Introduction}

Heavy electron materials manifest a broad spectrum of correlated-electron phenomena, including long-range antiferromagnetism, quantum criticality, and unconventional superconductivity \cite{LohneysenQPTreview,StewartHFreview,Paschen2020}.    In these systems, the $f$ electrons tend to be more localized, but interact weakly with itinerant conduction electrons via a Kondo interaction.  This interaction can lead to long-range magnetic order via a Ruderman-Kittel-Kasuya-Yosida (RKKY) interaction between the $f$ electron moments, or delocalization of the $f$ electrons by hybridizing with the conduction electrons and formation of a narrow band with enhanced mass~\cite{doniach}.  Frequently, pressure tuning uncovers a regime of non-Fermi liquid behavior that may be associated with an underlying quantum phase transition \cite{ColemanHFdeath,Gegenwart2008}. Many of these materials have tetragonal lattices, such as the prototypical CeMIn$_5$ family with M = Co, Rh, or Ir \cite{tuson,Thompson2006,sidorov} or the ThCr$_2$Si$_2$ structure of YbRh$_2$Si$_2$ \cite{Schuberth2016,Prochaska2020}, CeCu$_2$Si$_2$ \cite{steglichdiscovery} or URu$_2$Si$_2$ \cite{MapleURu2Si2}. Less is known, however, about the behavior of systems with other crystal symmetries.

\begin{figure}
\centering
\includegraphics[width=\linewidth]{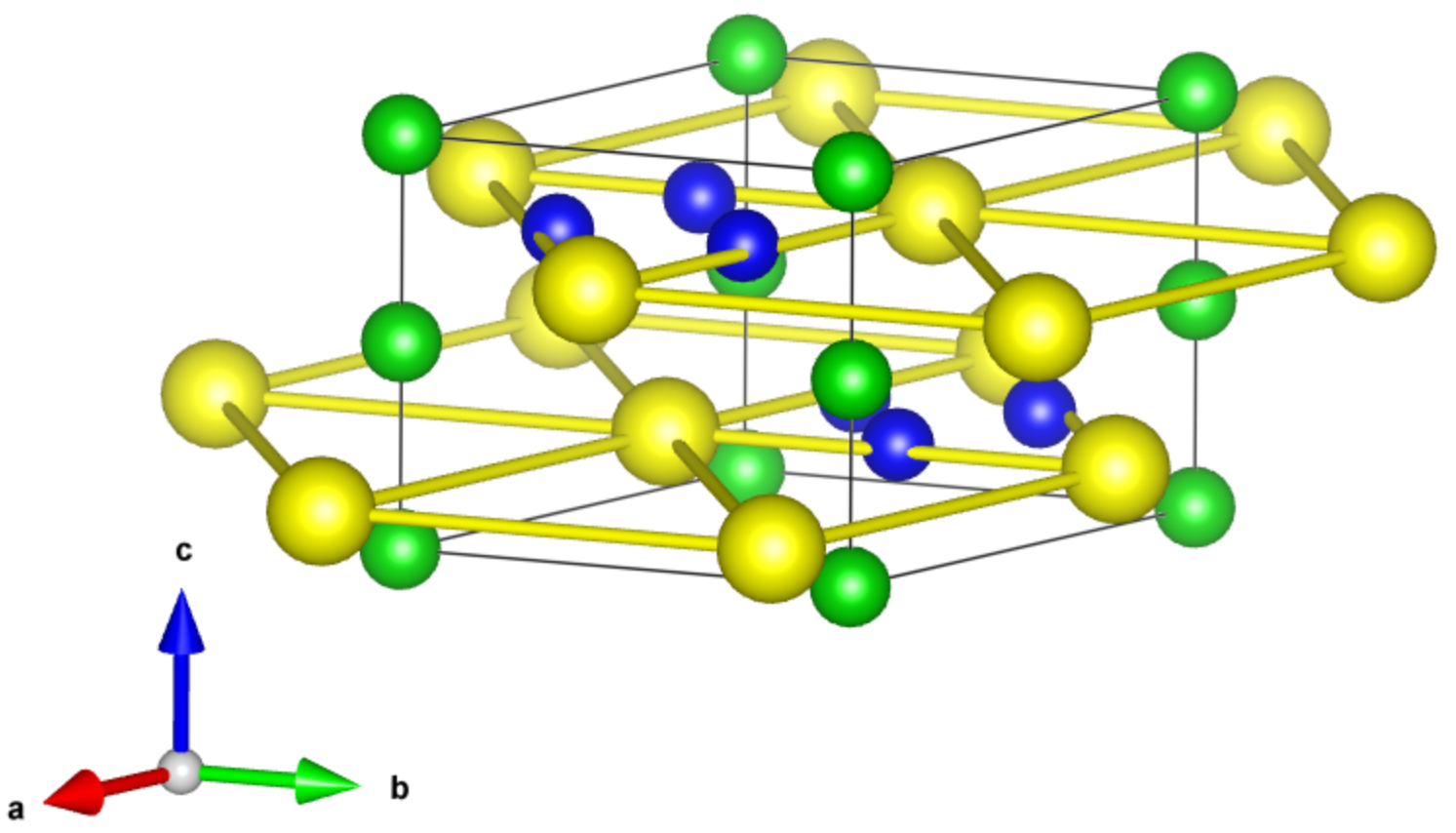}
\caption{\label{Fig:unitcell} Hexagonal crystal structure of CeVGe$_3$, with Ce in yellow, V in green, and Ge in blue. }
\end{figure}

CeVGe$_3$ crystallizes in the hexagonal $P6_3/mmc$ space group with two Ce atoms per unit cell, where the structure can be viewed as sheets of Ce with AB-type stacking, as illustrated in Fig.~\ref{Fig:unitcell}~\cite{Bie2009}. The nearest neighbor Ce-Ce distance of 4.58 \AA\ is similar to that in the CeMIn$_5$ materials, hence naively one might expect to find similar Kondo lattice physics. Indeed, CeVGe$_3$ exhibits antiferromagnetism below $T_N =5.5$\,K \cite{Inamdar2014}.  In the paramagnetic phase, the magnetic susceptibility reveals an effective moment that is close to that expected for Ce$^{3+}$ ($2.54\mu_B$), and the heat capacity exhibits a large peak at $T_N$. However, the entropy reaches only $0.5R\ln 2$  at 7\,K, suggesting a significant Kondo interaction.  Moreover, the contribution of the $4f$ orbitals to the resistivity indicates broad peaks at 30\,K and 50\,K along the $a$ and $c$ directions, indicative of the onset of Kondo lattice coherence \cite{NakatsujiFisk}.  It is interesting to note that the hexagonal sheet structure of this material gives rise to a Dirac dispersion at the $K$ points in momentum space.  Kondo physics is not expected to emerge in a Dirac semimetal when the Fermi level is precisely at the Dirac point, but can be present if the Dirac point lies below \cite{Mitchell2015}.

Little is known about the magnetic structure in CeVGe$_3$. Bulk measurements of the magnetization have revealed a Curie-Weiss temperature $\theta_\mathrm{CW}$ that is significantly higher than $T_N$, suggesting the presence of frustration \cite{Inamdar2014}.  On the other hand, crystalline electric field (CEF) and Kondo interactions can significantly alter the low temperature magnetic susceptibility \cite{CEF115study}.  A recent study of the magnetic response suggests that the CEF ground state in this material is $\Gamma_6$, corresponding to $|J_z=\pm 1/2\rangle$ \cite{Jin2022}. The measured magnetization anisotropy indicates that the easy axis lies in the basal plane, and the anisotropy agrees with the calculated $g$-factor for this state.  Moreover, the  first excited state doublet lies at $\Delta = 137$\,K, well above the Kondo lattice coherence temperature.  In-plane field-dependent measurements revealed a temperature-dependent metamagnetic transition around 2.5\,T \cite{Inamdar2014}.

In order to better understand the nature of the long-range order in this material, we have conducted detailed nuclear magnetic resonance (NMR), neutron scattering studies, and transport measurements on a single crystal and powder of CeVGe$_3$.  Our results reveal a helical structure in zero field, in which the Ce moments lie in the basal plane and rotate along the $c$-axis.  In-plane fields destabilize this structure, suppressing $T_N$ and giving rise to a first-order metamagnetic transition for fields $\gtrsim 2.5$\,T. The NMR spectra in this field-induced phase are consistent with a structure with partial polarization along the applied field and commensurate antiferromagnetism along the $c$-axis.  In the paramagnetic phase, the NMR Knight shift reveals anomalous behavior below a coherence temperature $T^*\sim 15$ K, consistent with the onset of Kondo lattice behavior, and the spin lattice relaxation rate indicates significant antiferromagnetic fluctuations.   These results indicate that although it has a hexagonal rather than a tetragonal lattice structure, CeVGe$_3$ exhibits behavior analogous to the archetypical heavy fermion CeRhIn$_5$.

\section{Kondo Physics}

\subsection{NMR Knight Shift}

\begin{figure}
\centering
\includegraphics[width=\linewidth]{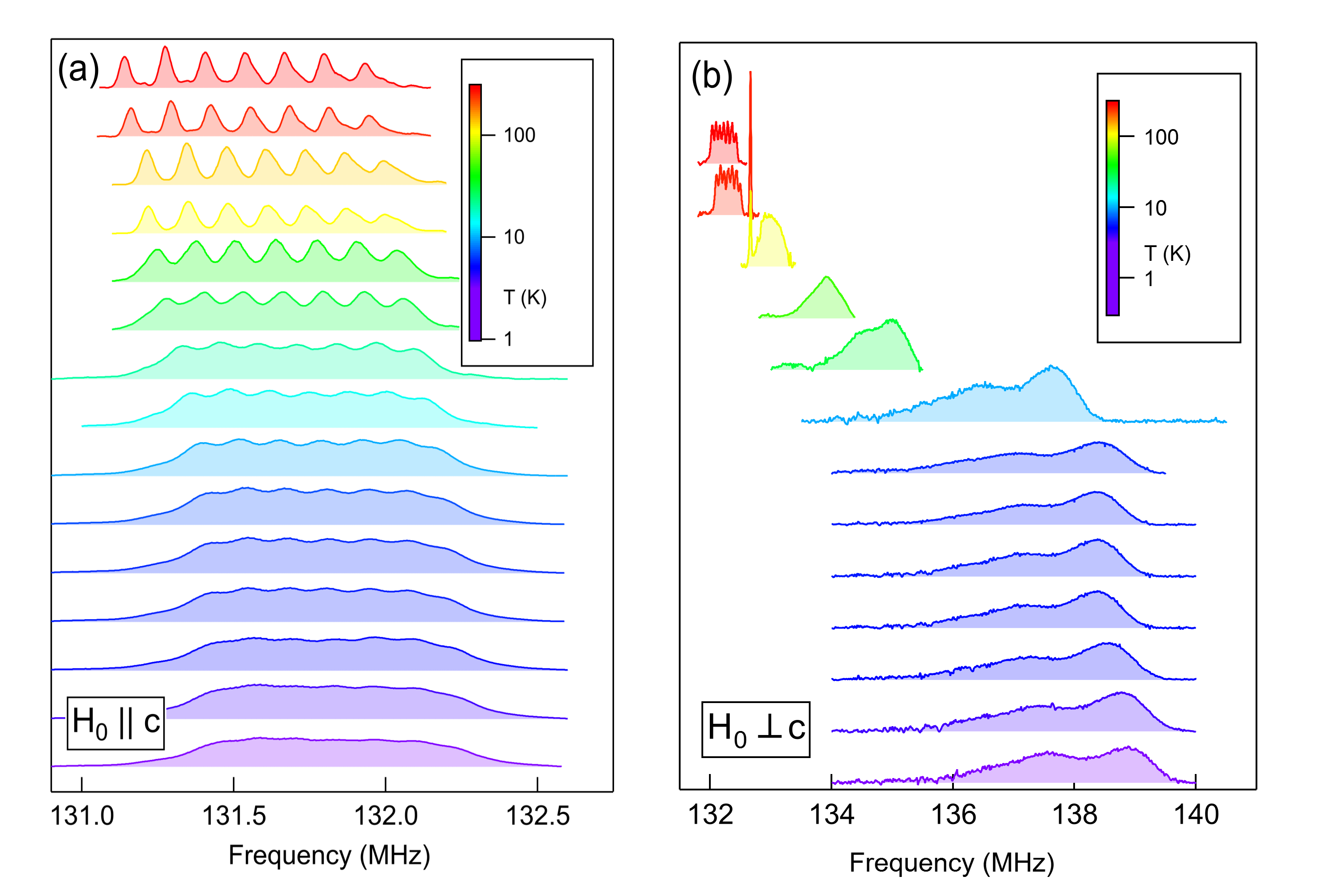}
\caption{\label{Fig:spectra} $^{51}$V  NMR spectra for (a) $\mathbf{H_0} \parallel [001]$ and (b) $\mathbf{H_0} \perp [001]$ as a function of temperature in an external field of 11.7285 T. }
\end{figure}

\begin{figure}
\centering
\includegraphics[width=\linewidth]{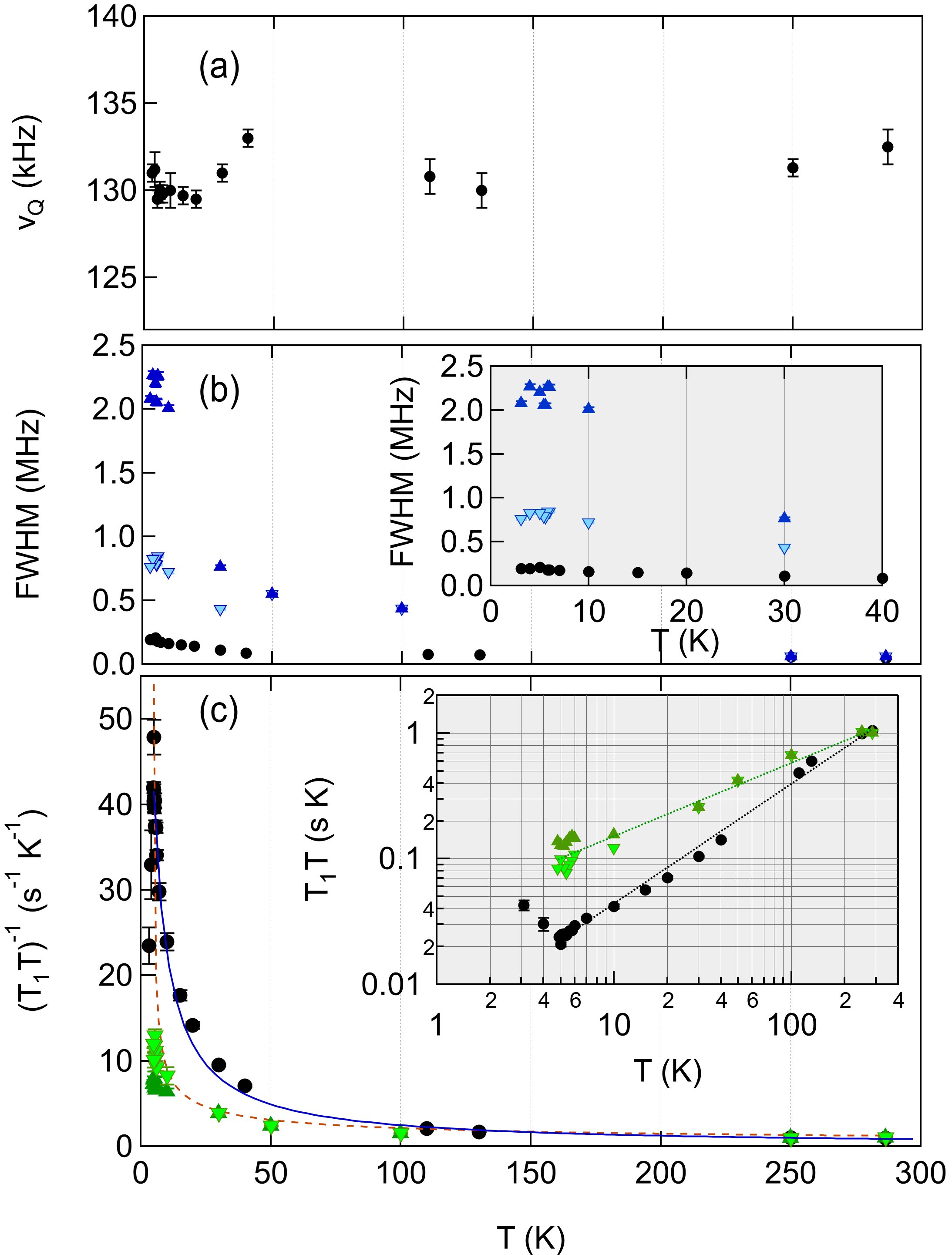}
\caption{\label{Fig:T1summary} (a) Electric field gradient, $\nu_{zz}$, versus temperature. (b) Linewidth versus temperature for $\mathbf{H}_0 \parallel c$ ($\bullet$) and $\mathbf{H}_0 \perp c$ ($\triangledown$, $\blacktriangle$). The inset focuses on the low temperature behavior. (c) $(T_1T)^{-1}$ versus temperature (same symbols as for (b)).   The solid line is a fit to a Curie-Weiss form, and the dashed line is a fit to the SCR expression, as discussed in the text. INSET: $T_1T$ versus temperature for both field directions.  The dotted lines are fits to a power law with exponent $0.95\pm0.01$ for $\mathbf{H}_0\parallel~c$. and $0.58\pm0.03$ for $\mathbf{H}_0\perp~c$.}
\end{figure}

\begin{figure}
\centering
\includegraphics[width=\linewidth]{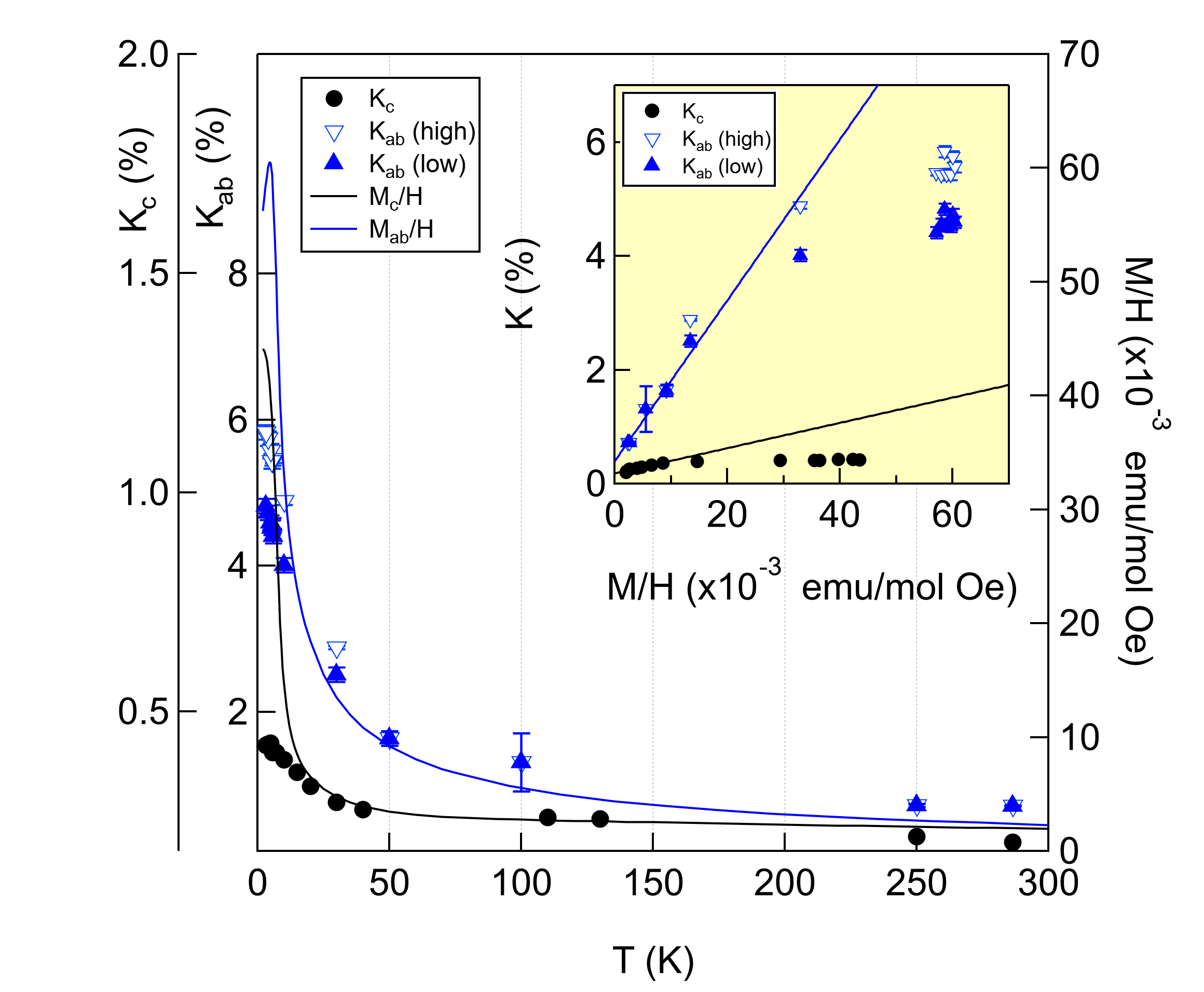}
\caption{\label{Fig:Knightshift} Knight shifts, $K_c$ ($\bullet$) and $K_{ab}$ ($\triangledown$, $\blacktriangle$) as a function of temperature compared with the bulk susceptibility, $M/H$ (solid lines), measured at 1 T. INSET: The Knight shift versus bulk susceptibility for both directions.  The solid lines are fits as described in the text.}
\end{figure}

A single crystal of CeVGe$_3$  with dimensions $5\times2\times 2$ mm$^3$ was secured in an NMR coil placed in an external field of  11.7 T. Spectra were acquired for fields both parallel and perpendicular to the $c$-axis by sweeping frequency as a function of temperature, as shown in Fig. \ref{Fig:spectra}. There is a single V site per unit cell, located centrosymmetrically between the Ce sheets. Because $^{51}$V is spin $I=7/2$, there are seven transitions at frequencies given by $\nu_n=\gamma H_0(1+K)+n \nu_q$, where $\gamma=11.193$ MHz/T is the gyromagnetic ratio, $K$ is the magnetic shift, $n=-3,\cdots,+3,$ and $\nu_q$ is the quadrupolar shift that arises due to the electric field gradient (EFG) tensor. The crystal has axial symmetry, hence the EFG tensor can be fully represented by a single quantity, $\nu_{zz} = (eQ/84h)\partial^{2}V/\partial z_{\alpha}^2$,
where $Q=5.2\times 10^{-30}{\rm m}^2$ is the quadrupolar moment of the $^{51}$V, and the quadrupolar shift $\nu_q = \nu_{zz}(3\cos^2\theta - 1)/2$ where $\theta$ is the angle between $\mathbf{H}_0$ and the $c$-axis.  The spectra were fit to a sum of Gaussians or Lorentzians to extract $K$, $\nu_{zz}$ and linewidths, as shown in Figs. \ref{Fig:T1summary}(a,b) and \ref{Fig:Knightshift}. In both cases, the spectra shift upwards in frequency with decreasing temperature, and the quadrupolar splitting between the peaks remains temperature independent. For $\mathbf{H}\perp~c$, the spectra broaden by an order of magnitude below $\sim 105$\,K, such that the quadrupolar splitting gets washed out.  Upon further cooling, the spectra exhibit two broad peaks. The shifts of these two peaks are labelled $K_{ab}$ (high/low) and shown in Fig.~\ref{Fig:Knightshift}. The strong temperature dependence of the shift and the broadening, accompanied by the lack of any change in the EFG, reflect properties of the magnetism, rather than any inhomogeneous broadening due to a distribution of local EFGs.

The temperature dependence of the  Knight shift tracks the bulk susceptibility for temperatures $T \gtrsim 30$ K.    This behavior arises due to the hyperfine coupling between the nuclear and electron spins: $\mathcal{H}_{hyp} = \hat{\mathbf{I}}\cdot\mathbb{A}\cdot\mathbf{S}_c + \hat{\mathbf{I}}\cdot\mathbb{B}\cdot\mathbf{S}_f$, where $\mathbf{S}_{c,f}$ correspond to the conduction or $f$-electron spins, and $\mathbb{A}$ and $\mathbb{B}$ correspond to the hyperfine tensors to these two spin species \cite{Curro2004,ShirerPNAS2012,jiang14}.  The Knight shift  and bulk susceptibility are given by:
\begin{eqnarray}
    K &=& A\chi_{cc} + (A+B)\chi_{cf} + B\chi_{ff}\\
    \chi &=& \chi_{cc} + 2\chi_{cf} + \chi_{ff}
\end{eqnarray}
where $\chi_{ij}=\langle S_{i}S_{j}\rangle$ and $(i,j) = (c,f)$.  For sufficiently high temperatures, the local moment susceptibility dominates, and $\chi\approx \chi_{ff}$. In this case $K = K_0 + B\chi$, and the transferred hyperfine coupling to the local moments is given by the slope in a plot of $K$ versus $\chi = M/H$, as shown in the inset of Fig. \ref{Fig:Knightshift}. We find that $B_{\parallel} = 0.4\pm 0.1$ kOe/$\mu_B$ and $B_{\perp} = 1.3\pm 0.2$ kOe/$\mu_B$.

These couplings presumably arise due to the overlap of the wavefunction of the Ce $4f$ and V $3d$ and $4s$ orbitals. In the CeMIn$_5$ system, the transferred hyperfine couplings are determined by the hybridization between the CEF ground state of the $4f$ wavefunction and the neighboring ligand sites \cite{Menegasso2021}.
Recently, a study of the magnetic properties of CeVGe$_3$ \cite{Jin2022} indicated that the ground state of the $4f$ is the $\Gamma_6$ manifold, with $|J_z = \pm1/2\rangle$, in contrast to the CeMIn$_5$ system \cite{Willers_2015}.  In this case, the $4f$ orbital is extended spatially along the $c$-axis, and may have a small overlap with the $V$ site. This fact may explain why the transferred hyperfine coupling in CeVGe$_3$ is much smaller than in CeRhIn$_5$, despite the fact that the gyromagnetic ratio of $^{51}$V and $^{115}$In are similar in magnitude \cite{Curro2000a,Curro2004}. Further studies of the hyperfine coupling in the Ce(V,Ti)Ge$_3$ may reveal dramatic changes, as the ground state wavefunction is expected to changes from $\Gamma_6$ to $\Gamma_7$ as a function of Ti doping~\cite{Jin2022}.

\begin{figure}
\centering
\includegraphics[width=\linewidth]{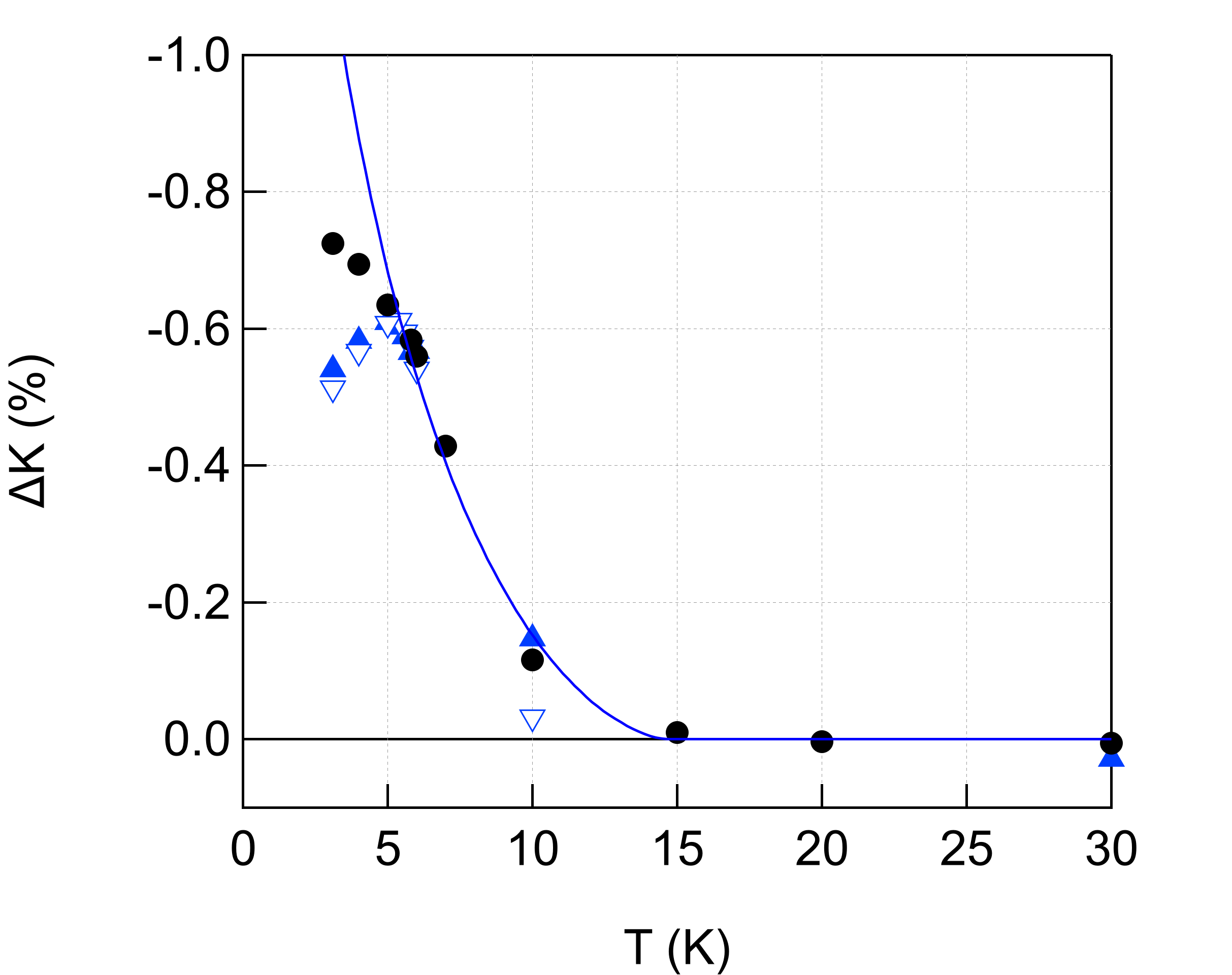}
\caption{\label{Fig:Kcf} $K_{cf}$ versus temperature for $\mathbf{H}_0\parallel~c$ ($\bullet$) and $\mathbf{H}_0\perp~c$ ($\triangledown$, $\blacktriangle$). The solid line is a fit as described in the text. For the high (low) shift in the perpendicular direction, the data have been scaled by a factor of 0.14 (0.18).}
\end{figure}

Below $T\approx 20$\,K, the Knight shift deviates from the bulk susceptibility, as shown in Fig. \ref{Fig:Knightshift}.  This behavior reflects the increase of $\chi_{cf}$ below the Kondo lattice coherence scale \cite{YangPinesPNAS2012,jiang14}.  In this case, we can construct the quantity $K_{cf} = K - B\chi \propto \chi_{cf} + \chi_{cc}$, which quantifies the growth of the heavy electrons.  This quantity is shown in Fig. \ref{Fig:Kcf} for both directions.  The data exhibit similar behavior for $T\gtrsim 5$\,K. The temperature dependence is expected to be described by:
\begin{equation}
K_{cf}(T) = K_{cf}^0(1-T/T^*)^{3/2}[1+\log(T^*\!/T)]
\label{eqn:YangPines}
\end{equation}
where $T^*$ is the coherence temperature and $K_{cf}^0$ is a temperature independent constant that depends on the undetermined on-site hyperfine coupling $A$ \cite{YangDavidPRL}.  The solid line in Fig. \ref{Fig:Kcf} is a fit to this expression for $T> 5$ K, which yields $T^*= 15\pm 1$ K.  Below $5$ K, this scaling breaks down such that $K_{cf}$ is less than the expected value in both directions. In fact, $K_{cf}$ actually decreases for the perpendicular direction. Similar behavior, known as relocalization,  has been observed in other heavy electron antiferromagnets, including CeRhIn$_5$ \cite{ShirerPNAS2012} and CePt$_2$In$_7$ \cite{Warren2010}. In this case, the physical picture is that the growth of the heavy electron fluid is interrupted by the growth of critical antiferromagnetic correlations as the temperature decreases towards $T_N$.

\subsection{Spin Lattice Relaxation Rate}

The spin lattice relaxation rate, \slrr, was measured by inversion recovery at the central transition ($I_z = +1/2 \leftrightarrow -1/2$). The magnetization recovery was fit to the standard expression for a spin 7/2 nucleus:
$M(t)\!=\!M_0\left(1-2f\sum_n A_n e^{-\alpha_n t/T_1}\right)$
where $M_0$ is the equilibrium nuclear magnetization, $f$ is the inversion fraction, $A_1 = 1225/1716$, $A_2 = 75/364$, $A_3 = 3/44$, $A_4= 1/84$, $\alpha_1 = 28$, $\alpha_2 = 15$, $\alpha_3  = 6$, and $\alpha_4 = 1$.  There was no evidence of any stretched relaxation.  For $\mathbf{H}_0\perp ~c$, \slrr\ was measured at both broad peaks at low temperatures where the quadrupolar splitting was no longer resolved. The temperature dependence of $(T_1T)^{-1}$ is shown in Fig. \ref{Fig:T1summary}(c).  For both directions this quantity diverges strongly with reduced temperature, and exhibits a peak at 5 K, which is consistent with bulk measurements of $T_N$.

The spin lattice relaxation rate probes the dynamical part of the spin susceptibility through the relation:
\begin{equation}
	\label{eqn:dynamical_susceptibility}
	\left(\frac{1}{T_1T}\right)_{\alpha} = \gamma^2 k_B \lim_{\omega \rightarrow 0} \sum\limits_{\mathbf{q},\beta\neq\alpha} \mathcal{F}_{\alpha\beta}(\mathbf{q}) \frac{\textrm{Im}\chi_{\alpha\beta}(\mathbf{q},\omega)}{\hslash \omega},
\end{equation}
where $\mathcal{F}_{\alpha\beta}(\mathbf{q})$ are form factors that depend on the hyperfine coupling tensor, $\chi_{\alpha\beta}(\mathbf{q},\omega)$ is  the dynamical magnetic susceptibility, and $\alpha,\beta = \left\{ x,y,z \right\}$ \cite{MoriyaT1formula}.  The large peak in $(T_1T)^{-1}$ reflects the slowing down of critical fluctuations near $T_N$.

According to the self-consistent renormalized
(SCR) theory of spin fluctuations for weak itinerant
antiferromagnets $(T_1T)^{-1} \propto \chi_Q(T)^n$, where $n=1/2$ for 3D fluctuations and $n=1$ for 2D fluctuations \cite{Moriya1974}. The dotted and solid lines in Fig. \ref{Fig:T1summary}(c) are fits to $n=1/2$ and $n=1$ with $\chi_Q(T) = C/(T+\theta_\mathrm{CW})$, where $\theta_\mathrm{CW} = -4.8 \pm 0.1$ K for $n=1/2$, and $\theta_\mathrm{CW} = 1.0\pm 0.1$K for $n=1$.  The solid line provides a much better fit to the data for $\mathbf{H}_0\parallel~c$, suggesting that the fluctuations are predominantly 2D in nature for this direction.
The inset of Fig. \ref{Fig:T1summary}(c) shows $T_1T$ versus temperature for both directions, as well as power law fits.  For the parallel direction, the best fit yields an exponent $n =0.95\pm0.01$, but for the perpendicular direction, the best fit yields $n=0.59\pm0.03$.  This result suggests that for the perpendicular field, the fluctuations exhibit 3D character in contrast to the parallel direction.  This behavior may be related to the presence of a different magnetic structure above the metamagnetic transition in the perpendicular direction, as discussed below. The related heavy fermion antiferromagnet CeRhIn$_5$ exhibits more 3D spin fluctuations \cite{Curro2000a,Mito2001}.  On the other hand, $(T_1T)^{-1}$ in the the heavy fermion superconductor CeCoIn$_5$ exhibits an unusual $T^{1/4}$ behavior that has been associated with proximity to an antiferromagnetic quantum critical point \cite{ZhengIrCo115NMR2003}. The superconductor CeIrIn$_5$ exhibits behavior that is in some sense a hybrid between 2D and 3D, with $(T+\theta)^{3/4}$ that has been associated with the layered structure \cite{ZhengIr115PRL}.

\section{Magnetic Ordering}

\subsection{Neutron Scattering}

\begin{figure*}
\centering
\includegraphics[width=\linewidth]{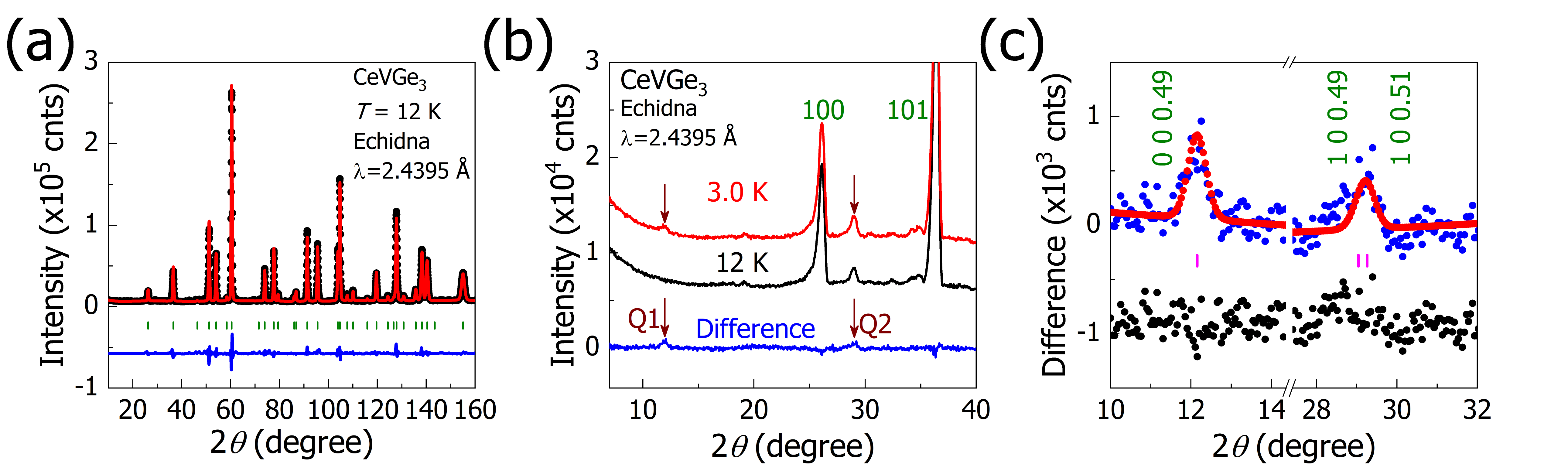}
\caption{\label{Fig:ND1} (a) Neutron powder diffraction pattern (black circles) at 12 K, along with the corresponding Rietveld refinement (red line). The green bar represents the allowed positions of nuclear reflections. The negligible difference pattern (blue line) between the nuclear contribution and the refined pattern confirms the accuracy and reliability of the crystal structure model.  (b) Neutron powder diffraction pattern at 3.0 K (red line) and 12 K (black line), respectively, offset for clarity. The difference pattern (blue line) indicates the presence of magnetic reflections, as indicated by the arrows. (c) The difference pattern (blue circles) between 3.0 K and 12 K, along with the simulated pattern of in-plane helical magnetic structure (red circles). The pink bars indicate the allowed magnetic reflection positions. The negligible difference (black circles) between the observed magnetic contribution and the simulated pattern confirms the validity of the estimated moment size.}
\end{figure*}

The Kondo effect can cause a reduction in the magnetic moment of a material. To investigate the size of the ordered moment in CeVGe$_3$, neutron powder diffraction experiments were conducted below and above $T_N$. Figure \ref{Fig:ND1}(a) displays the neutron powder diffraction pattern obtained at 12 K (above $T_N$), along with the corresponding Rietveld refinement. The refined parameter and reliable factors are summarized in Table \ref{tab:table1}. The result indicates that the crystal structure at 12 K is the same as that investigated at room temperature \cite{Jin2022}, which is depicted in Fig. \ref{Fig:unitcell}. Additionally, Fig. \ref{Fig:ND1}(b) provides an enlarged view of the low-angle region, measured at temperatures of 3.0 K and 12 K. The difference pattern reveals the emergence of two distinct magnetic scattering intensities at positions $|\mathbf{Q}_1|$ and $|\mathbf{Q}_2|$. It is not possible to index these reflections with rational numbers, suggesting the presence of a potential incommensurate magnetic ground state in CeVGe$_3$. Assuming that $|\mathbf{Q}_1|$ occurs along the high symmetry direction, the magnetic propagation vector ($\vec{k}$) can be
indexed as $(0.46, 0, 0)$ or $(0.33, 0.33, 0)$ in the basal plane (in units of the reciprocal lattice vectors). If $|\mathbf{Q}_1|$ aligns with the $c$ axis, the estimated $\vec{k}$  value would be approximately $(0, 0, 0.48)$.

To determine the exact $\vec{k}$, single crystal neutron diffraction was performed with the crystal aligned with the $HK0$ plane as the scattering plane. One-dimensional (1D) reciprocal-space scans along $(H, 0, 0)$, $(-H, H, 0)$ and $({H},{H}, 0)$ at base temperature 2.7 K did not reveal any magnetic reflection. Additionally, a two-dimensional (2D) reciprocal-space mesh scan at 2.7 K was conducted to detect satellite magnetic reflections, where $H\neq K$ in the basal plane. However, no additional scattering intensity was observed (Fig. \ref{Fig:ND4} in Appendix C), strongly suggesting that $\vec{k}$ does not predominantly lie  within the basal plane. Alternatively, magnetic reflections may be present in regions outside of the basal plane.

\begin{figure*}
\centering
\includegraphics[width=0.8\linewidth]{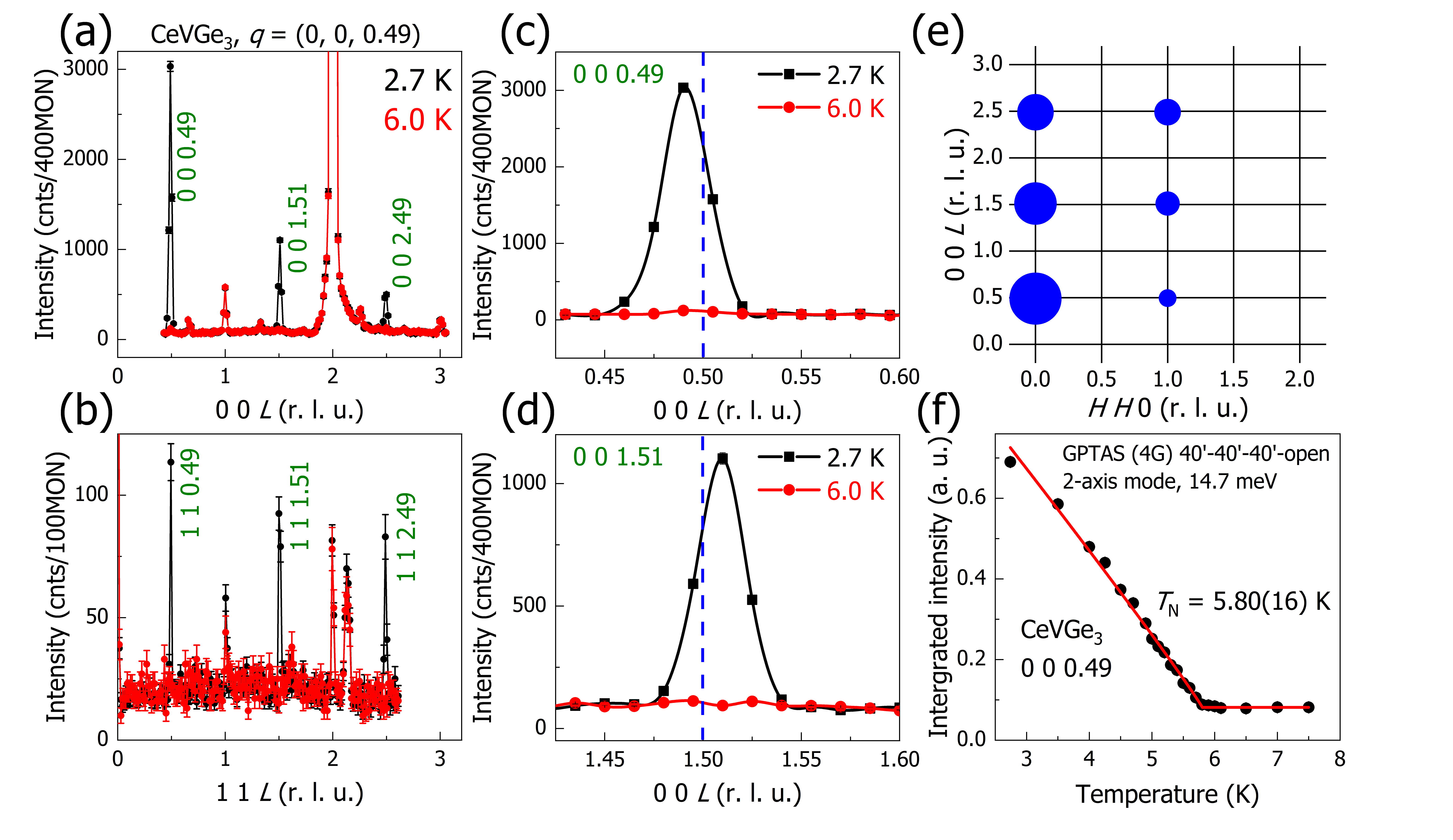}
\caption{\label{fig:ND2} The raw dataset of the reciprocal space $L$-scan along the (a) $(0, 0, {L})$ and (b) $(1, 1, {L})$ direction at 2.7 K and 6.0 K. The selected reciprocal space $L$-scan along the $(0, 0, {L})$ direction with an enlarged view at (c) (0, 0, 0.49) and (d) (0, 0, 1.51). The blue dashed lines indicate the locations at (0, 0, 0.5) and (0, 0, 1.5), respectively. (e) The distribution of the squared magnetic structure factor at 2.7 K (below \emph{T}\textsubscript{N}) in  the ${HHL}$ plane strongly suggests that magnetic moment lies in the basal plane. (f) The temperature dependence of the integrated intensity of the 0 0 0.49 magnetic reflection. The magnetic transition temperature is estimated to be 5.80(16) K.}
\end{figure*}

The  single crystal was then rotated by 90 degrees from the ${HK}0$ plane to the ${HHL}$ plane as the new scattering plane. The 1D reciprocal-space scans along the $(0, 0, {L})$ and $(1, 1, {L})$ directions at 2.7 K and 6.0 K were plotted and shown in Fig. \ref{fig:ND2}(a,b). No magnetic reflections were observed on the top of the nuclear or forbidden nuclear positions in the ${HHL}$ plane. On the other hand, satellite magnetic reflections were observed between these nuclear peaks. Figures \ref{fig:ND2}(c,d) display an enlarged view of the selected magnetic reflections at 2.7 K and 6.0 K. Their peak positions were determined to be $(0, 0, 0.49)$ and $(0, 0, 1.51)$. Figure \ref{fig:ND2}(e) illustrates the distribution of squared magnetic structure factors in ${HHL}$ plane. In this case, satellite magnetic reflections (Fig. \ref{fig:ND2}(e)) were successfully observed at positions $(0, 0, 0.49)$, $(0, 0, 1.51)$, $(0,0, 2.49)$, $(1, 1, 0.49)$, $(1, 1, 1.51)$, and $(1, 1, 2.49)$. It should be noted that other than the allowed nuclear reflections, some peaks were present along $(0, 0, {L})$ and $(1, 1, {L})$ directions. All of them are temperature independent and are not intrinsic. The reflections at
$(0, 0, 1)$, $(0, 0, 3)$, and $(1, 1, 1)$ are likely due to the higher-harmonic contributions, while those at $(0, 0, 2.26)$ and $(1, 1, 2.13)$ are from the Al sample holder. We also find some intensity at $(0, 0, 0.655)$, $(0, 0, 1.33)$, and $(0, 0, 2.665)$, which could arise from an unknown impurity phase included in the crystal. None of these reflections affect the analysis discussed below.

Based on the data from the ${HK0}$ and the ${HHL}$ planes, we conclude that the ground state exhibits a single-${k}$ incommensurate magnetic (ICM) structure with $\vec{k}=(0, 0, 0.49)$. The temperature dependence of the integrated intensity of the most intense magnetic reflection $(0, 0, 0.49)$ is depicted in Fig. \ref{fig:ND2}(f). From the fit to the power-law function with a constant background, we estimate $T_N=5.80(16)$ K, which is consistent with the magnetization measurements \cite{Inamdar2014, Jin2022}.

To analyze incommensurate magnetic (ICM) structures, we employed the magnetic representation method. The allowed irreducible representations (IRs) for the Ce ($2{d}$) site within the $P6_{3}/mmc$ space group with ${k} = (0, 0, 0.49)$ are summarized in Table \ref{tab:table2} in Appendix C. Examination of the basis vectors associated with each IR reveals that IR2 and IR5 are 1D IRs, while IR3 and IR6 belong to 2D IRs. Models based on IR2 and IR5 were discarded because they resulted in allowed magnetic structures where the Ce layers align ferromagnetically along \emph{c} axis and stack antiferromagnetically. These results are inconsistent with the magnetization data \cite{Inamdar2014, Jin2022}. Additionally, it was observed that the simulated magnetic intensities did not align well with the observed intensities. For instance, magnetic reflections along the $00{L}$ direction were not expected due to the polarization factor. On the other hand, IR3 and IR6 correspond to in-plane helical structures. In the case of IR3, the calculated magnetic structure factor at $(0, 0, 0.49)$ is zero, while the most intense magnetic structure factor appears at $(0, 0, 0.51)$. These results are in direct contradiction with the experimental observations. Only IR6 reproduces the positions and the intensities of the observed magnetic reflections.

\begin{figure}
\centering
\includegraphics[width=0.8\linewidth]{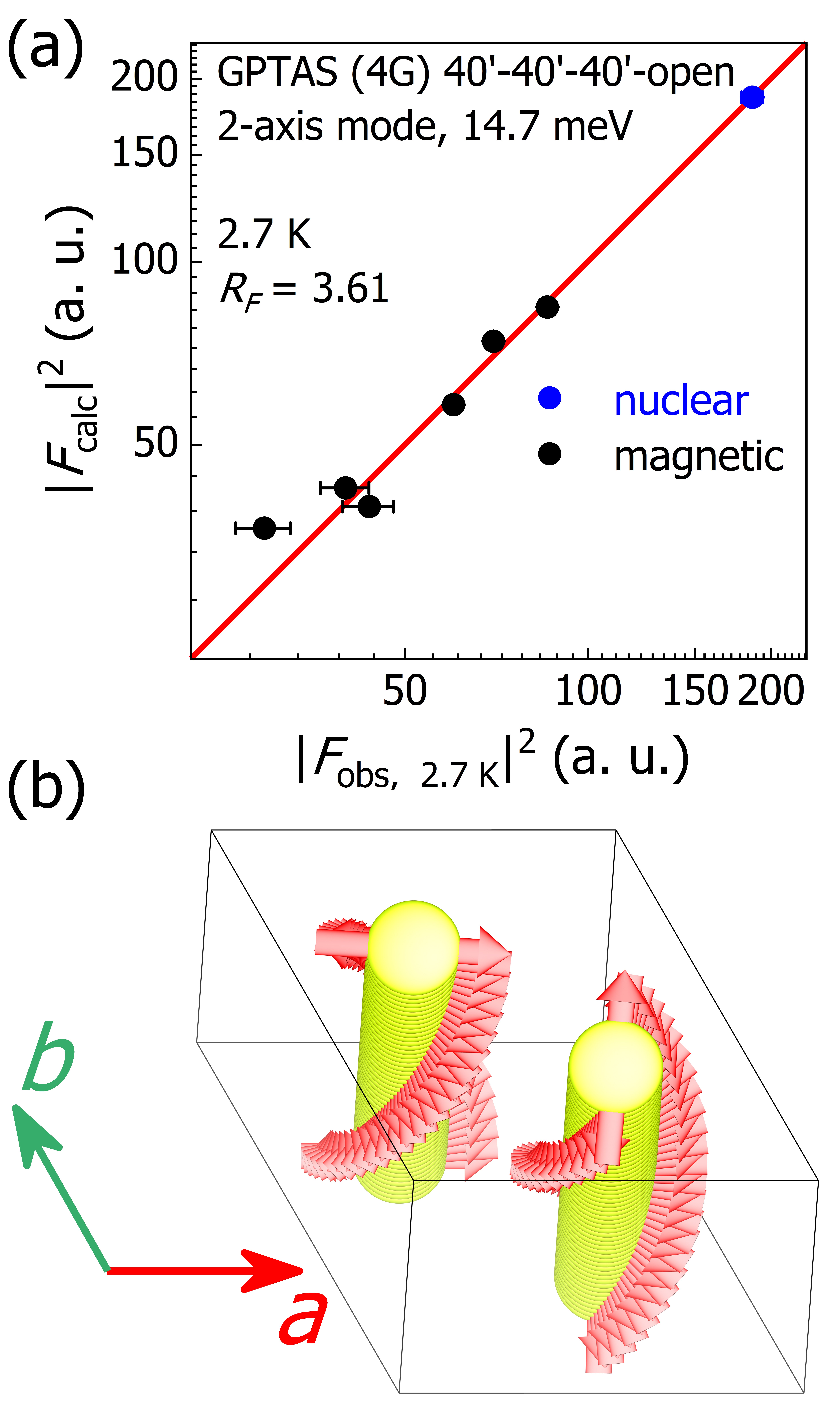}
\caption{\label{Fig:NDspiral} (a) The calculated magnetic structure factor from the refinement versus the observed magnetic structure factor obtained at 2.7 K.  (b) The proposed in-plane helical magnetic structure at zero-field obtained through refinement. The yellow circles are the two Ce sites per unit cell in adjacent layers, and the red arrows indicate the magnetic moments.}
\end{figure}

The magnetic structure factors obtained from the single crystal data measured at 2.7 K were utilized for a least-square refinement employing the IR6 model. To simplify the refinement process and reduce the number of parameters, the atomic position and lattice constants were fixed to those obtained from powder diffraction at 12 K. In addition, a simplifying assumption was made that the moment size of all Ce atoms is the same. This reduces the number of refinement parameters from two to one, corresponding to the moment size. The helical structure with a fixed moment size is represented by a single BV, where BV1 represents a counter-clockwise rotating helical structure and BV2 represents a clockwise helical structure. Since they could not be distinguished through refinement, the coefficient of BV1 was used as an adjustable parameter. The calculated and observed magnetic structure factors were compared in Fig. \ref{Fig:NDspiral}(a). A scale factor was determined from the 110 nuclear reflection. Despite having only a single adjustable parameter, the six observed magnetic reflections were well reproduced. The estimated moment size at 2.7 K was 0.49(1) $\mu_B$.

The magnetic structure was further confirmed through neutron powder diffraction, as shown in Fig. \ref{Fig:ND1}(c). The peak at $|\mathbf{Q}_1|$ was indexed as 0 0 0.49, while the peak at $|\mathbf{Q}_2|$ was a superposition of 1 0 0.49 and 1 0 0.51 reflections. A moment size of 0.38(1) $\mu_B$ at 3.0 K was estimated from the Rietveld refinements using the difference pattern. The red circles in Fig. \ref{Fig:ND1}(c) represent the helical structure with a fixed moment size, and accurately reproduce the intensity of the two peaks. The estimated moment sizes from both data sets are very close, with any small differences likely attributable to the slight variation in sample temperature.

\subsection{Hyperfine Field}

\begin{figure}
\centering
\includegraphics[width=\linewidth]{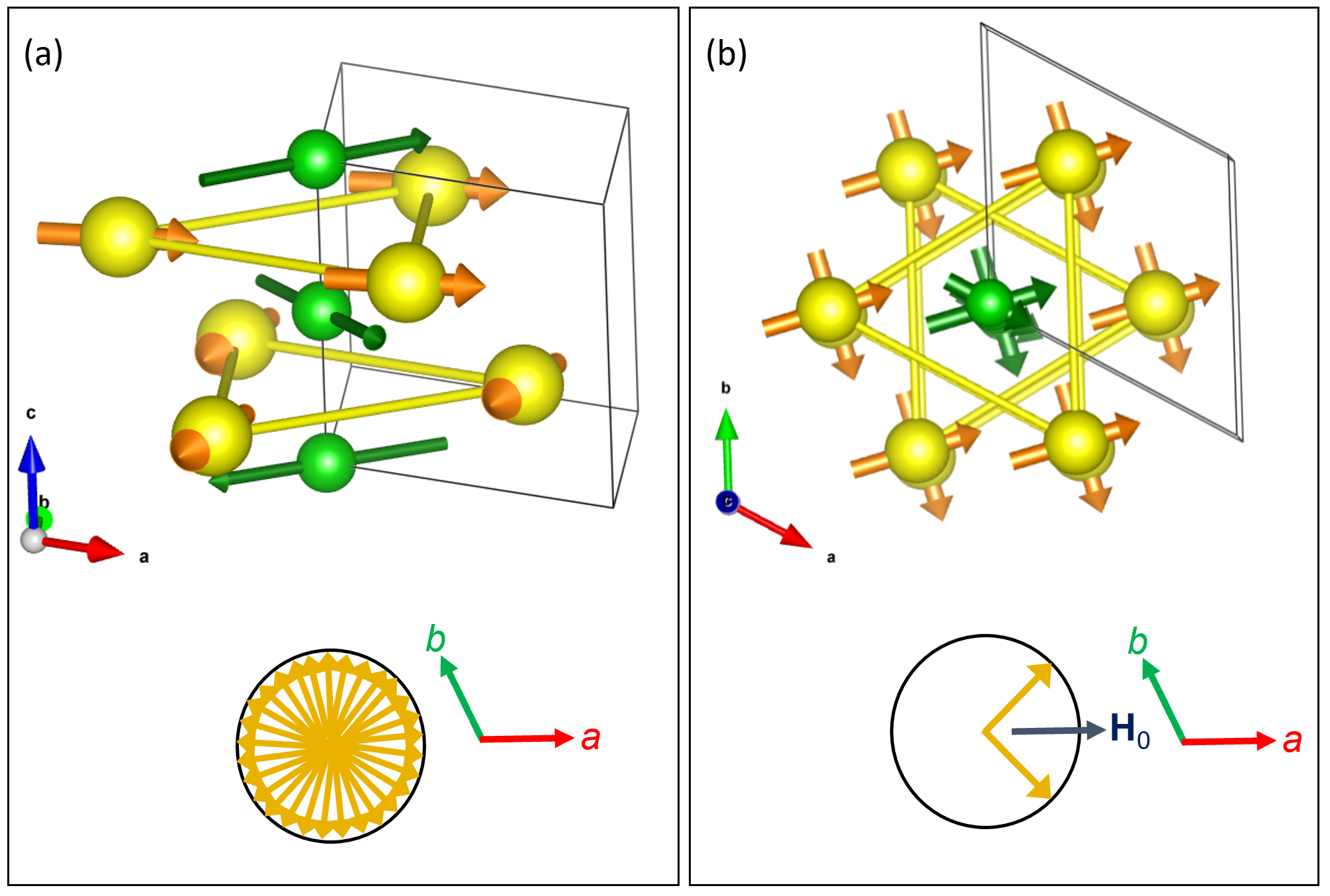}
\caption{\label{Fig:HFfield} (a) Magnetic structure and hyperfine fields in the low-field helicoidal phase. Orange arrows indicate ordered moments on Ce sites (yellow), and green arrows indicate hyperfine field at the V sites (green). The lower illustration indicates the projection of the Ce moments in the $ab$ plane. (b) Magnetic structure and hyperfine fields in the field-induced phase.  The external field here is directed along the crystalline $a$-axis, indicated by the red arrow. The lower illustration indicates the projection of the Ce moments in the $ab$ plane relative to an in-plane field, $\mathbf{H}_0$.  }
\end{figure}

In the magnetically ordered state, the NMR resonance will be affected by a static hyperfine field given by:
$\mathbf{H}_{hf} =\sum_{i=1}^{6}\mathbb{B}_i\cdot\mathbf{m}(\mathbf{r}_i)$, where the $\mathbb{B}_i$ are the transferred hyperfine couplings to the 6 n.n. Ce sites, as discussed in  Appendix B. For a helical structure of the Ce ordered moments with wavevector $Q_z$, the internal field rotates in the $ab$ plane with constant magnitude, as illustrated in Fig. \ref{Fig:HFfield}(a), and is given by $H_{hf} = 6B_{\perp} m_0\left|\cos\left({2\pi k_z z_0}/{2c}\right)\right|$, where $z_0 = c/2$ is the spacing between the Ce layers. Using the measured hyperfine couplings, and the neutron results $m_0 = 0.38 \mu_B$ and $k_z = 0.49$, we estimate $H_{hf} = 0.21$ T.

In the presence of an external field, $\mathbf{H}_0$, the NMR resonance frequency is given by $f= \gamma\left| \mathbf{H}_0 + \mathbf{H}_{hf}\right|$.  If $\mathbf{H}_0~||~c$, then the resonance frequency in the ordered state is given by $f=\gamma(H_0^2 + H_{hf}^2)^{1/2}$, and is homogeneous. For the applied field of 11.7 T, the resonance is shifted upwards by only 22 kHz. As seen in Fig. \ref{Fig:spectra}(a) and \ref{Fig:T1summary}(b), the spectrum for $\mathbf{H}_0~||~c$ changes little through $T_N$, but already has a linewidth of $\sim200$ kHz. It is likely that inhomogeneous demagnetization fields broaden the resonance sufficiently that the subtle shift  due to the hyperfine field is washed out \cite{Vinograd2022}. Thus the NMR response for this direction is consistent with the neutron scattering results.

\subsection{Metamagnetic Transition}

\begin{figure}
\centering
\includegraphics[width=\linewidth]{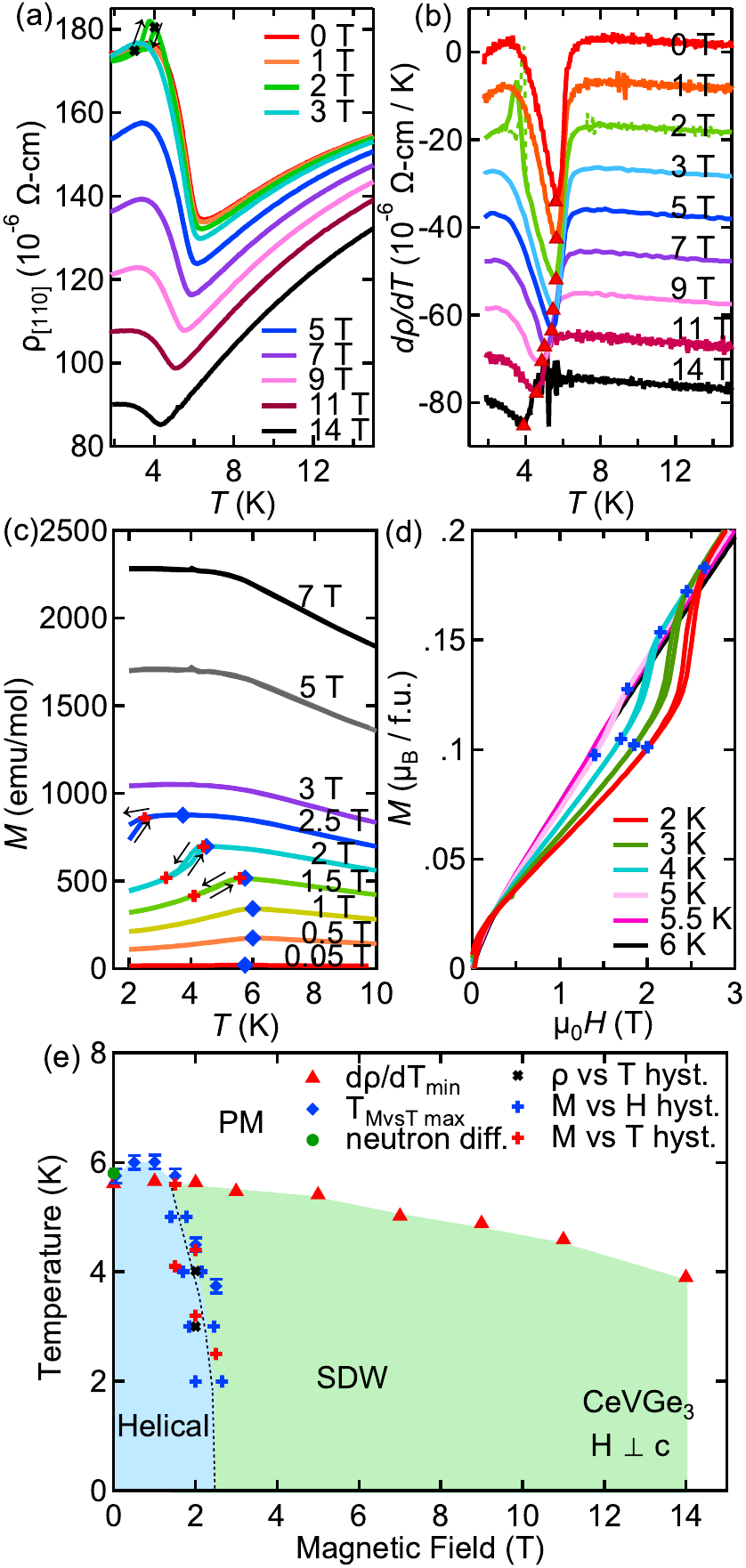}
\caption{\label{Fig:phasediagram} (a) Temperature dependence of the in-plane resistivity $\rho(T)$ of a CeVGe$_3$ single crystal under various field up to 14\,T. (b) Evolution of the temperature derivative of the resistivity at the low temperature under different fields. The data are vertically offset by 10\,$\mu\Omega\,\mathrm{cm}\,\mathrm{K}^{-1}$ to reduce overlap. (c) Temperature dependence and (d) magnetic field dependence of in-plane magnetization of CeVGe$_3$. (e) $H-T$ phase diagram of CeVGe$_3$ with in-plane applied fields. The criteria to map out the phase diagram are marked in (a)-(d). Two symbols mark the beginning and end of the hysteresis.}
\end{figure}

Previous work uncovered a metamagnetic transition for fields perpendicular to the $c$-axis \cite{Inamdar2014}. Helical magnetic order is known to be unstable to the presence of an in-plane field \cite{Nagamiya1962}, and has been observed in CeRhIn$_5$ \cite{baoCeRhIn5INS,Raymond2007,Mishra2021}.  To determine the magnetic phase diagram of CeVGe$_3$, we carried out resistivity measurements up to 14\,T and magnetization measurements up to 7\,T with the field applied perpendicular to the c-axis as shown in Fig.~\ref{Fig:phasediagram}.

The temperature-dependent resistivity at zero field shows a sharp upturn near $T_N$ (Fig.~\ref{Fig:phasediagram}(a)), possibly due to the fact that the incommensurate helical order is accompanied by Fermi surface nesting. The peak positions of the derivative of resistivity are marked by red triangles to track the field dependence of $T_{N}$. As shown in Fig.~\ref{Fig:phasediagram}(b), $T_{N}$ is monotonically suppressed with increasing field. As can be seen in Fig.~\ref{Fig:phasediagram}(a), the resistivity at 2\,K is decreasing with increasing field, indicating that the magnetic scattering due to the helical order starts to be suppressed above 3\,T.

As shown in Fig.~\ref{Fig:phasediagram}(c), the anomaly corresponding to the antiferromagnetic transition in the temperature dependence of the magnetization changes: the maximum (labeled as blue diamonds) slightly increases with the field below 1\,T, and is then suppressed near 2.5\,T. Above 3\,T, there is no obvious anomaly above 2\,K.

The spin-flop transition
is also detected as a field-induced jump in magnetization with some hysteresis, as shown in Fig.~\ref{Fig:phasediagram}(d). The magnetic structure above the spin-flop transition has yet to be determined, so we generally label it as another spin-density wave due to the possible nesting observed in resistivity. The phase diagram for in-plane fields is thus summarized in Fig. \ref{Fig:phasediagram}(e).

In the helical phase at zero field, the angle between the Ce moments in adjacent planes is 88.2$^{\circ}$.  This value is close to $90^{\circ}$, so is likely that the most stable structure in the high field SDW phase is commensurate \cite{Nagamiya1962}.  In this case, the moments remain in the basal plane, with partial polarization along $\mathbf{H}_0$ (e.g. $\mathbf{m}(\mathbf{r}) = {m}_0\{\cos\alpha,\sin\alpha\cos(\pi z/c - \pi/4),0  \}$ for field along $[100]$). This $++--$ structure corresponds to an up-up-down-down sequence of magnetic moments when moving along the $c$-axis. This arrangement of the proposed magnetic structure can be further interpreted as a composite of two magnetic components, where the superposition results in a ferromagnetic (FM) component with $\vec{k} = \{ 0,0,0\}$ and an antiferromagnetic (AFM) component with $\vec{k} = \{ 0,0,0.5\}$. Here $\alpha$ characterizes the angle between the moments and $\mathbf{H}_0$, and for sufficiently high fields $\alpha\rightarrow 0$ corresponds to a fully polarized state.  A similar structure was found in CeRhIn$_5$ \cite{Raymond2007}, however CeRhIn$_5$ also exhibits a third phase with an incommensurate structure just below $T_N$. We find no evidence for such a region in the phase diagram of CeVGe$_3$.

The NMR measurements for $\mathbf{H}_0\perp c$ were carried out at 11.7\,T in the metamagnetic phase.  The spectra are clearly inconsistent with a helical magnetic structure, which would generate a broad double-peaked spectrum centered at the Larmor frequency with two ``horns" at $\pm \gamma H_{hf}\approx 3.7$\,MHz.  Rather the spectrum is shifted well above the Larmor frequency, as observed in Fig.~\ref{Fig:spectra}(b).

On the other hand, our observations are consistent with the proposed $++--$ structure, as illustrated in Fig. \ref{Fig:HFfield}(b).  In this case, the hyperfine field has two possible values, depending upon whether the V site lies between Ce layers with the same spin directions ($++$ or $--$, case A) or with alternating directions ($+-$ or $-+$, case B):
\begin{eqnarray}
\mathbf{H}_{hf}= \begin{cases}
\pm 6 B_{\perp} m_0\{\cos\alpha,\sin\alpha,0 \}\textrm{ case A}\\
6 B_{\perp} m_0 \{\cos\alpha,0,0 \} \textrm{\hspace{0.72cm} case B}\end{cases}.
\end{eqnarray}
This scenario thus offers a natural explanation for the presence of the upper and lower sites in the NMR spectra shown in Fig. \ref{Fig:spectra}(b).  The frequency shift for the two cases are:  $\Delta f_A = \sqrt{f_0^2 + (6 \gamma B_{\perp} m_0)^2 + 12 \gamma f_0 B_{\perp} m_0 \cos\alpha} - f_0$ and $\Delta f_B = 6 \gamma B_{\perp} m_0\cos\alpha$, where $f_0 = \gamma H_0$. These equations can be inverted to extract $m_0$ and $\alpha$.  At the lowest temperatures, we find $m_0 = 2.3\mu_B$ and $\alpha = 72^{\circ}$ in an applied field of 11.7\,T. Note that this value of $m_0$ is larger than that expected for a $\Gamma_6$ ground state (1.3$\mu_B$) \cite{Jin2022}. It is possible that this result could be modified by mixing of excited crystal field levels by the applied field, however.  The strong enhancement of the susceptibility and Knight shift in this field direction with decreasing temperature seen in Fig. \ref{Fig:Knightshift} reflects the partial polarization of the Ce spins prior to the onset of long-range order below $T_N$, as captured by the angle $\alpha$.

\section{Discussion}

Helical magnetic structures generally arise in layered structures with frustrated in-plane and out-of-plane interactions \cite{Nagamiya1962}.  For example, in a model with ferromagnetic in-plane interaction, $J_0<0$, as well as  both $J_1$ between nearest-neighbor layers, and $J_2$ between next-nearest-neighbor layers, the ground state magnetic structure is an incommensurate spiral with wavevector $k_z = \cos^{-1}(-J_1/4J_2)/\pi$. We therefore find $|J_1/J_2|  \approx 0.13$ for CeVGe$_3$, with $J_{1}>0$ and $J_2<0$. Moreover, the metamagnetic field $H_m$ should be related to $J_2$ as $H_m\approx 1.8 J_2/\mu_B$, suggesting that $J_2\sim -1 $ K \cite{Nagamiya1962}.  On the other hand, the same model predicts that the spins would be fully polarized at $\approx 2 H_m \sim 5$ T, but we find no evidence of any other phase transition up to 14 T. Of course, the magnetic interactions in Ce-based heavy fermion is more complex than this simple model.  Spatially-varying couplings that can give rise to spiral order can arise naturally via the RKKY interaction, but depend  on the details of the Fermi surface and the Kondo interaction \cite{NSRh1152014}.

As seen in Fig. \ref{Fig:spectra},  the spectra for $\mathbf{H}_0 \perp c$ reveal two separate peaks emerging below $T\approx 30$ K, well above $T_N$. This is surprising because in principle the system remains paramagnetic.  The origin of the two peaks in this phase remains unclear, but potentially could be related to a distribution of demagnetization fields \cite{Vinograd2022}.

As a heavy fermion compound that exhibits Kondo lattice coherence and helical antiferromagnetic order at ambient pressure, CeVGe$_3$ exhibits similarities to the archetypical heavy fermion system, CeRhIn$_5$. On the other hand, there are important differences: CeVGe$_3$ has hexagonal rather than tetragonal structure, the ratio of the excited state doublet energy to the Kondo coherence scale,  $\Delta/T^*$, is approximately 9.0 in CeVGe$_3$ versus 4.5 in CeRhIn$_5$ \cite{ShirerPNAS2012}, and the CEF ground state manifold may be $\Gamma_6$ rather than $\Gamma_7$.  The orbitals of the latter are extended more in the plane, whereas those of the latter lie more along the $c-$axis \cite{WillersCEF115s}. The spatial form of these Ce orbitals can play a significant role in the anisotropic Kondo interactions, and may give rise to the different properties of the CeMIn$_5$ series as $M$ changes due to slight alterations in the CEF interactions \cite{HauleCeIrIn5,Willers_2015}.  It would therefore be of interest to examine how the ground state of CeVGe$_3$ evolves both as a function of doping and hydrostatic pressure. CeRhIn$_5$ becomes superconducting under approximately 2 GPa.  It would also be valuable to examine the phase diagram of CeVGe$_3$
at high magnetic fields.  It is also important to investigate the field dependence of $T_N$ to higher fields in CeVGe$_3$. A recent specific heat study up to 36 T in CeRhIn$_5$ identified a new field-induced phase for fields along the $c$-axis, possibly to a new incommensurate phase \cite{Mishra2021}.

\begin{acknowledgements}

Work at UC Davis was supported by Work at UC Davis was supported by the NSF under Grant No.  DMR-2210613, as well as the UC Laboratory Fees Research Program (LFR-20-653926). We acknowledge support from the Physics Liquid Helium Laboratory fund. The study at Tohoku University was supported by Grants-in-Aid for Early Career Scientists (Grant No. 20K14395), Scientific Research on Innovative Areas (Grants No. 19H05824), Scientific Research (A) (Grant No. 22H00101), Challenging Research (Exploratory) (Grant No. 19K21839),  the Fund for the Promotion of Joint International Research (Fostering Joint International Research; Grant No. 19KK0069) from the Japan Society for the Promotion of Science, and the CORE Laboratory Research Program “Dynamic Alliance for Open Innovation Bridging Human, Environment and Materials” of the Network Joint Research Center for Materials and Device. The experiments at JRR-3 (proposal No. 22900) and the travel expenses for the neutron scattering experiment at ANSTO were partly supported by the General User Program for Neutron Scattering Experiments, Institute for Solid State Physics, University of Tokyo.

\end{acknowledgements}

\appendix

\section{Synthesis method}
The synthesis of the polycrystalline CeVGe\textsubscript{3} sample was prepared using the arc-melting method, followed by annealing the
as-grown ingot sealed in a quartz tube at 850 \textsuperscript{o}C for 15 days. Single crystals of CeVGe$_3$ were synthesized via the self-flux method. The starting materials [Ce pieces (Ames Lab), V pieces (etched with nitric acid), Ge lumps (6N)] were initially arc-melted to ensure a homogeneous mixture. The initial composition of elements is $\mathrm{Ce}:\mathrm{V}:\mathrm{Ge} = 4:1:19$. To synthesize large single crystals that are suitable for NMR and neutron scattering study, we modify the temperature profile compared with the previous reports~\cite{Inamdar2014, Jin2022}. The arc-melted mixture was placed in a $2$\,mL Canfield Crucible Set~\cite{Canfield:uq}, and sealed in a fused silica ampoule in a partial pressure of argon. The sealed ampoule was placed in a furnace where it was held at \SI{1200}{\celsius} for 10 hours, and slowly cooled to \SI{860}{\celsius} over 210 hours. At \SI{860}{\celsius}, the ampoule was removed from the furnace and quickly centrifuged to separate the single crystals from the molten flux. The largest single crystal we managed to grow is about 70\,mg.

\section{Hyperfine coupling}

There are six nearest neighbor Ce ions to each V in CeVGe$_3$, and these are arranged in a hexagonal array such that three are above the V and three are below.
The general form of the hyperfine coupling between a moment $\mathbf{m}(\mathbf{r})$ on the Ce located at $\mathbf{r}$ and the nuclear spin $\mathbf{I}$ on the V site is:
\begin{equation}
  \mathcal{H}_{hf} = \gamma \hbar \mathbf{I}\cdot\sum_{i=1}^{6}\mathbb{B}_i\cdot\mathbf{m}(\mathbf{r}_i)
\end{equation}
where the hyperfine coupling matrices are:
\begin{equation}
  \mathbb{B}_i = \left(
                   \begin{array}{ccc}
                     B_{i,aa} & B_{i,ab} & B_{i,ac} \\
                     B_{i,ab} & B_{i,bb} & B_{i,bc} \\
                     B_{i,ac} & B_{i,bc} & B_{i,cc} \\
                   \end{array}
                 \right)
\end{equation}
for each of the six nearest neighbors.  These are related to one another by various symmetries.  For example, $\{\mathbb{B}_1,\mathbb{B}_3,\mathbb{B}_5\}$ and $\{\mathbb{B}_2,\mathbb{B}_4,\mathbb{B}_6\}$ rotate into one another by $\frac{2\pi}{3}$ rotations about $z$, and  that the second set is related the first by both a $\frac{\pi}{3}$ rotation around $z$ and a mirror reflection across the $ab$ plane.  As a result, there are only 6 independent quantities $B_{\alpha\beta}$.

We can compute the Knight shift tensor by assuming that $\langle \mathbf{m}\rangle = \chi\mathbf{H}_0$ is uniform. This leads a diagonal tensor with elements $K_{aa} = K_{bb} =6B_{\perp}\chi_{aa}$ and $K_{cc} = 3B_{\parallel}\chi_{cc}$, where $B_{\perp} =(B_{aa} + B_{bb})/2$, and $B_{\parallel} = B_{cc}$.  We can measure these quantities experimentally by comparing $K_{\alpha\alpha}$ and $\chi_{\alpha\alpha}$ to determine the components $B_{\perp}$ and $B_{\parallel}$ as discussed in the text.

Below $T_N$ in the helical phase the ordered moments are given by: $\mathbf{m}(\mathbf{r}) = m_0\{ \cos(2\pi k_z z/c),\, \sin(2\pi k_z z/c),\, 0\}$, the internal hyperfine field lies in the $ab$ plane and is given by  $H_{hf} = H^0_{hf}\{ \cos(2\pi k_z z'/c),\, \sin(2\pi k_z z'/c),\, 0\}$, where $H^0_{hf} = 6B_{\perp} m_0\left|\cos\left({2\pi k_z z_0/c}/{2}\right)\right|$ and $z' = z-c/2$.

\section{Neutron scattering}

\begin{table*}
\caption{\label{tab:table1} The refined crystallographic data, numbers of nuclear reflections, and reliable factors of CeVGe$_3$ at 12 K. The isotropic displacement $B_{iso}$ parameter for three atoms is fixed to 0.1 to reduce the number fitting parameters. The profile $R_p = \sum_i |y_{obs}-y_{calc} |/\sum_i y_{obs}$, and the weighted profile $R_{wp} = \left[\sum_i w_i |y_{obs}-y_{calc}|^2/\sum_i w_i y_{obs}^2\right]^{1/2}$; The expected $R_{exp} = \left[(N-P+C)/\sum w_i y_{obs}^2\right]^{1/2}$.}
\begin{center}
    CeVGe$_3$\\
Hexagonal $P6_3/mmc$, (No. 194), $T$ = 12 K (Echidna)\\
$a = 6.1988(1)$ \AA, $b = 6.1988(1)$ \AA, $c = 5.6470(1)$ \AA\\
$\alpha = 90^{\circ}$, $\beta = 90^{\circ}$, $\gamma = 120^{\circ}$\\
Number of fitting parameters 21\\
Number of nuclear reflections 30
\end{center}
\begin{ruledtabular}
    \begin{tabular}{cccccc}
 Atom & Site & $x$ & $y$ & $z$ & $B_{iso}$ (\AA$^2$) \\ \hline
        Ce & 2$d$ & 1/3 & 2/3 & 3/4 & 0.1 \\
        V & 2$a$ & 0 & 0 & 0 & 0.1 \\
        Ge & 6$h$ & 0.1954 (2) & 0.3908 (3) & 1/4 & 0.1
    \end{tabular}
\end{ruledtabular}
\begin{center}
    $R_p = 12.9$\%, $R_{wp} = 16.6$\%, $R_{exp} = 2.22$\%, $\chi^2 = 55.8$
\end{center}
\end{table*}

\begin{table*}
\caption{\label{tab:table2} The allowed irreducible representations for the Ce $2d$ sites within the $P6_3/mmc$ space group, considering the magnetic modulation vector $\vec{k} = (0, 0, 0.49)$. The notation for the moment direction is given as $(m_a, m_b, m_c)$, where $m_c$ is parallel to the sixfold axis and $m_a$ and $m_b$ are along the  $a$ and $b$ axes, respectively, forming an angle of 120 degrees. Here $a=\cos{(\pi\cdot0.49)}$ and $b=\sin{(\pi\cdot0.49)}$}
\begin{ruledtabular}
\begin{tabular}{lcc}
        Allowed IRs & ($1/3$, $2/3$, $3/4$) & ($2/3$, $1/3$, $1/4$) \\
        \hline
        IR2:BV1 & $(0, 0, 1)$ & $(0, 0, a+bi)$ \\
        IR3:BV1 & $(1-(1/\sqrt{3})i, -(2/\sqrt{3})i, 0)$ & $(-a-(1/\sqrt{3})b-(b-(1/\sqrt{3})a)i , -(2/\sqrt{3})b+(2/\sqrt{3})ai, 0)$ \\
        IR3:BV2 & $(-(2/\sqrt{3})i,1-(1/\sqrt{3})i,0)$ & $(-(2/\sqrt{3})b+(2/\sqrt{3})ai,-a-(1/\sqrt{3})b-(b-(1/\sqrt{3})a)i,0)$ \\
        IR5:BV1 & $(0, 0, 1)$ & $(0, 0, -a-bi)$ \\
        IR6:BV1 & $(1-(1/\sqrt{3})i, -(2/\sqrt{3})i, 0)$ & $(a+(1/\sqrt{3})b+(b-(1/\sqrt{3})a)i , (2/\sqrt{3})b-(2/\sqrt{3})ai, 0)$ \\
        IR6:BV2 & $(-(2/\sqrt{3})i,1-(1/\sqrt{3})i,0)$ & $((2/\sqrt{3})b-(2/\sqrt{3})ai,a+(1/\sqrt{3})b+(b-(1/\sqrt{3})a)i,0)$ \\
    \end{tabular}
\end{ruledtabular}
\end{table*}

\begin{table*}
\caption{\label{tab:table3} The fixed parameters of CeVGe$_3$ are used to estimate the magnetic moment size at 3.0 K.}
\begin{center}
    CeVGe$_3$\\
Hexagonal $P6_3/mmc$, (No. 194), $T$ = 3 K (Echidna)\\
$a = 6.19865(1)$ \AA, $b = 6.19865(1)$ \AA, $c = 5.6468(1)$ \AA\\
$\alpha = 90^{\circ}$, $\beta = 90^{\circ}$, $\gamma = 120^{\circ}$\\
\end{center}
\begin{ruledtabular}
    \begin{tabular}{cccccc}
 Atom & Site & $x$ & $y$ & $z$ & $B_{iso}$ (\AA$^2$) \\ \hline
        Ce & 2$d$ & 1/3 & 2/3 & 3/4 & 0.1 \\
     \end{tabular}
\end{ruledtabular}
\begin{center}
Number of refined parameters: 1 \\
Number of magnetic reflections: 15\\
\end{center}
\begin{ruledtabular}
    \begin{tabular}{ccc}
Atom&	Site &	$|M|$ \\\hline
Ce &	2$d$ &	$0.38(3) \mu_B$
     \end{tabular}
\end{ruledtabular}
\end{table*}

Neutron powder diffraction experiments were carried out using the high-resolution powder diffractometer Echidna installed at the OPAL reactor, Australian Nuclear Science and Technology Organisation (ANSTO) with a neutron wavelength of 2.4395 Å selected by the Ge 331 reflection. The mass of the CeVGe3 powder sample used in the neutron powder diffraction experiment was approximately 4.42 g. For the single crystal neutron diffraction experiments, the GPTAS (4G) triple-axis thermal neutron spectrometer without the analyzer (double-axis mode) installed at JRR-3, Tokai, Japan was used with the incident neutron energy of 14.7 meV and vertically focusing PG 002 monochromator. An additional PG filter was inserted after monochromator to suppress the higher harmonic neutron contribution. The CeVGe$_3$ single crystal with dimensions of $3\times2\times 1$ mm$^3$ and a mass of approximately 30 mg was aligned with two scattering planes, ${HK0}$ and ${HHL}$, and then sealed in a standard aluminum sample can with $^{4}$He exchange gas. A Gifford-McMahon (GM) cooler, which was a $^{4}$He closed cycle refrigerator, was utilized to cool down the sample to a base temperature of 2.7 K. 1D reciprocal-space scans along the $(0, 0, {L})$ and $(1, 1, {L})$ directions were conducted to collect the integrated intensity to obtain the squared structure factor. The Lorentz factor was numerically corrected in the analysis.

\begin{figure}
\centering
\includegraphics[width=0.8\linewidth]{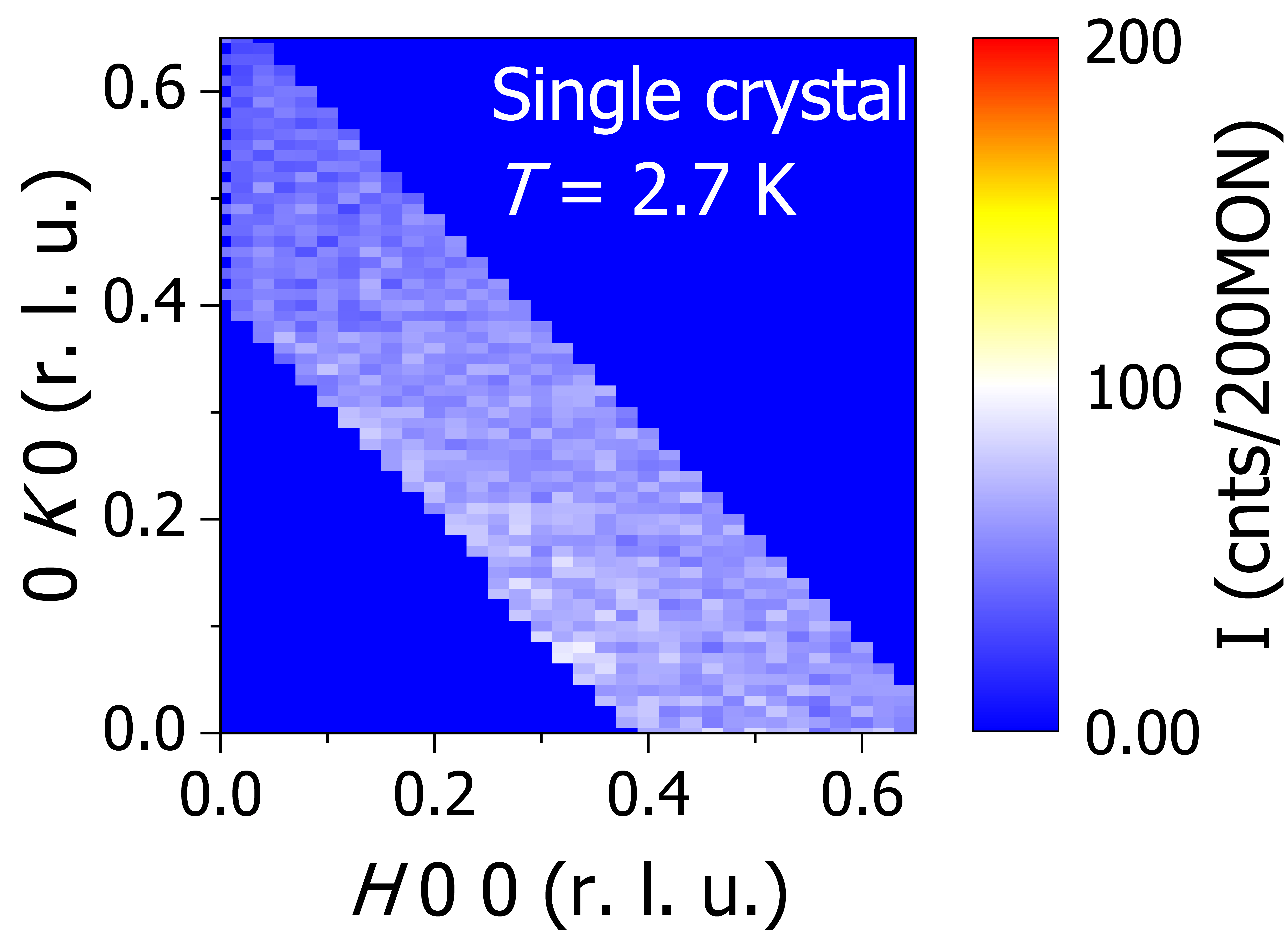}
\caption{\label{Fig:ND4} Two-dimensional mesh scan at 2.7 K, revealing the absence of any satellite magnetic reflections in the ${HK}0$ plane.}
\end{figure}

\clearpage

\bibliography{CeVGe3Bibliography}

\end{document}